\documentclass[useAMS,usenatbib]{mn2e}
\usepackage{rotating}
\usepackage{journals}

\def\approxgt{\ifmmode \rlap{$>$}{}_{{}_{{}_{\textstyle\sim}}} \else%
$\rlap{$>$}{}_{{}_{{}_{\textstyle\sim}}}$\fi} 
\def\approxlt{\ifmmode \rlap{$<$}{}_{{}_{{}_{\textstyle\sim}}} \else%
$\rlap{$<$}{}_{{}_{{}_{\textstyle\sim}}}$\fi}

\LARGE \normalsize \title[The neutron star soft X--ray transient XTE~J1709--267]{Optical
and X--ray observations of the neutron star soft X--ray transient XTE~J1709--267 }

\author[P.G. Jonker et al.]  {P.G.~Jonker$^{1,2}$\thanks{email :
    pjonker@cfa.harvard.edu}, D.K.~Galloway$^3$, J.E.~McClintock$^1$,
    M.~Buxton$^4$, M.~Garcia$^1$, S.~Murray$^1$\\
    $^1$Harvard--Smithsonian Center for Astrophysics, 60 Garden
    Street, Cambridge, MA 02138, Massachusetts, U.S.A.\\ $^2$Chandra
    Fellow\\ $^3$Massachusetts Institute of Technology, Center for
    Space Research, Cambridge, MA 02139, U.S.A.\\ $^4$Department of
    Astronomy, Yale University, 260 Whitney Avenue, New Haven, CT
    06520, U.S.A.}

\begin{document}

\maketitle

\begin{abstract} \noindent  In this Paper we report on the discovery of the
optical counterpart to the neutron star soft X--ray transient (SXT)
XTE~J1709--267 at an $R$--band magnitude of $R$=20.5$\pm$0.1 and
22.24$\pm$0.03, in outburst and quiescence, respectively. We further report the
detection of type~I X--ray bursts in {\it RXTE} data obtained during an
outburst of the source in 2002. These bursts show a precursor before the onset
of the main burst event, reminiscent of photospheric radius expansion bursts.
Sifting through the archival {\it RXTE} data for the burster 4U~1636--53 we
found a nearly identical burst with precursor in 4U~1636--53. A comparison of
this burst to true photospheric radius expansion bursts in 4U~1636--53 leads us
to conclude that these bursts--with--precursor do not reach the Eddington
limit. Nevertheless, from the burst properties we can derive that the distance
to XTE~J1709--267 is consistent with the distance of the Globular Cluster
NGC~6293. We further report on the analysis of a 22.7 ksec observation of
XTE~J1709--267 obtained with the {\it Chandra} satellite when the source was in
quiescence. We found that the source has a soft quiescent spectrum which can be
fit well by an absorbed black body or neutron star atmosphere model. A power
law contributes less than $\sim$20 per cent to the 0.5--10 keV unabsorbed flux
of $(1.0\pm0.3)\times10^{-13}$ erg cm$^{-2}$ s$^{-1}$. This flux is only
slightly lower than the flux measured right after the outburst in 2002. This is
in contrast to the recent findings for MXB~1659--29, where the quiescent source
flux decreased gradually by a factor of $\sim$7--9 over a period of 18 months.
Finally, we compared the fractional power--law contribution to the unabsorbed
0.5--10 keV luminosity for neutron star SXTs in quiescence for which the
distance is well--known. We find that the power--law contribution is low only
when the source quiescent luminosity is close to $\sim 1 - 2\times10^{33}$ erg
s$^{-1}$. Both at higher and lower values the power--law contribution to the
0.5--10 keV luminosity increases. We discuss how models for the quiescent
X--ray emission can explain these trends.

\end{abstract}

\begin{keywords} stars: individual: (XTE~J1709--267, 4U~1636--53) --- 
accretion: accretion discs --- binaries: general --- stars: neutron
stars --- X-rays: binaries
\end{keywords}

\section{Introduction}
\label{intro}

Low--mass X--ray binaries are binary systems in which a $\approxlt1
M_{\odot}$ star transfers matter to a neutron star or a black hole. A
large fraction of the low--mass X--ray binaries are transients; these
are called soft X--ray transients (SXTs). Although several of the
neutron star systems were already detected in quiescence with
Einstein, EXOSAT, ASCA, and ROSAT (e.g.~\citealt{1981BAAS...13..900P};
\citealt{1987A&A...182...47V}; \citealt{1994A&A...285..903V};
\citealt{1998PASJ...50..611A}; \citealt{1999ApJ...514..945R}), detailed
studies of these faint quiescent counterparts to neutron star
transients have only become possible with {\it Chandra} and XMM/Newton
(e.g.~\citealt{2001ApJ...560L.159W}, \citealt{2002ApJ...577..346R}).

Several mechanisms have been proposed to explain the observed X--ray luminosity
and spectra of quiescent neutron star SXTs. Accretion may be ongoing at a low
level producing a soft spectrum (\citealt{1995ApJ...439..849Z}). Several authors
have pointed out that the presence of a $\sim10^8$ Gauss magnetic field would
have a large influence on the accretion flow. The onset of a propeller or pulsar
wind mechanism (\citealt{1975A&A....39..185I}; \citealt{1986ApJ...308..669S};
\citealt{2000ApJ...541..849C}) has been proposed as an explanation of the hard
power--law spectral component. Although detailed theoretical model calculations
predicting the spectral shape are absent so far, \citet{1998ApJ...494L..71Z}
argue that in the propeller phase the spectrum will be hard.  In addition to
these two apparently mutually exclusive models, it is thought that a soft,
thermal spectral component with a luminosity of typically $10^{32-33} {\rm
erg\,s^{-1}}$ will be generated due to the fact that the neutron star crust and
core are heated via pycnonuclear reactions in the crust during the accretion
phase. The crust will thermally radiate in (soft) X--rays, cooling the neutron
star (e.g.~\citealt{1998ApJ...504L..95B}; \citealt{2001ApJ...548L.175C};
\citealt{2001MNRAS.325.1157U}). In this model the observed quiescent luminosity
can differ from outburst to outburst by a factor of 2--3 since the fraction of
hydrogen and helium left in the atmosphere after an outburst will vary from
outburst to outburst. This fraction influences the heat flux that flows from the
core to the surface (\citealt{2002ApJ...574..920B}). The quiescent luminosity is
also likely to vary somewhat from source to source since the neutron star core
and crust temperature depend on the mass accretion history of the source and the
cooling rate may depend among other things on the neutron star mass. Using
hydrogen neutron star atmosphere models to fit the soft part of the quiescent
spectrum, neutron star radii and temperatures can be determined.  The observed
values are in the range expected from neutron star theory. If it can be
established that the quiescent emission is indeed due to the hot neutron star
surface or core, these systems could provide a way to determine neutron star
radii. Together with information about the neutron star spin (e.g.~obtained
through burst oscillations [e.g.~\citealt{2003strohbild}] and/or pulsations
observed during outburst [e.g.~\citealt{2004wijn}]) and mass (see
\citealt{1999ApJ...512..288T}) this provides important information about the
behaviour of matter under physical conditions that are unattainable on Earth.

Recently, using the {\it Chandra} satellite we followed the neutron
star SXT RX~J170930.2--263927, also called XTE~J1709--267, towards
quiescence (\citealt{2003MNRAS.341..823J}). XTE~J1709--267 was detected
for the first time using ROSAT All Sky Survey observations performed
in 1990 (\citealt{1999A&A...349..389V}). The source was also detected by
ROSAT in 1992 (see \citealt{2001A&A...368..137V};
\citealt{2003MNRAS.341..823J}). Since then, the source has been detected
with RXTE three times; in 1997 (\citealt{1997IAUC.6543....2M}), in 2002
(\citealt{2003MNRAS.341..823J}), and in 2004
(\citealt{2004ATel..255....1M}). During the 1997 outburst
\citet{1998ApJ...508L.163C} found type~I bursts using the Wide Field
Cameras onboard the {\it BeppoSAX} satellite. Since the source is
located only 9--10 arcminutes away from the core of the Globular
Cluster NGC~6293 it has been speculated that XTE~J1709--267 is
associated with NGC~6293 (\citealt{2001A&A...368..137V};
\citealt{2003MNRAS.341..823J}).

In this Paper we present X--ray and optical observations of
XTE~J1709--267 in quiescence and outburst. A preliminary announcement
of the optical observations was already made in \citet{ateljonker}.

\section{Observations and analysis}
\subsection{Chandra}

We have observed the neutron star SXT XTE~J1709--267 during quiescence using the
ACIS--I CCDs operated in the very faint mode on board the {\it Chandra} satellite
(\citealt{2002PASP..114....1W}) for $\sim$25 ksec on May 12, 2003 (observation ID
3507). Due to the short deadtime introduced by reading out the CCDs the effective
on--source time was 22.7 ksec. The X--ray data were processed by the {\it Chandra}
X--ray Centre but we reprocessed the data starting with the level 1 products in order
to take full advantage of the newest available calibrations. We used the {\it CIAO}
software to reduce the data (version 3.0.2 and CALDB version 2.26). Events with ASCA
grades of 1, 5, 7, cosmic rays, hot pixels, and events close to CCD node boundaries
were rejected.  We searched the data for periods of enhanced background radiation but
none was present. Hence, all the data were used in our analysis.

We offset pointed the satellite with respect to the known accurate
coordinates of XTE~J1709--267 (see \citealt{2003MNRAS.341..823J}) in
order to put some of the Globular Cluster NGC~6293 in the
field--of--view.  We detected $>$15 sources, but the coordinates of
only one were consistent with those of XTE~J1709--267 (the analysis of
the other sources will be presented elsewhere). We detected 166 source
counts in 22.7 ksec. The spectrum of XTE~J1709--267 was extracted from
a circular region with a 5 arc second radius centred on the source
whereas the background spectrum was extracted from a circular region
with a radius of 5 arc seconds located 50 arc seconds East of the
source (3 background counts were detected in this region). We rebinned
the spectrum such that each bin contained at least 5 counts per
bin. Because of this low number of counts we used the CASH statistic
method in our spectral fitting to estimate the errors on the fitting
parameters (\citealt{1979ApJ...228..939C}). We only include photons with
energies above 0.3 and below 10 keV in our spectral analysis since the
ACIS timed exposure mode spectral response is not well calibrated
outside that range. In order to validate the CASH statistics we did
not subtract the background photons. These background photons all have
energies above 3 keV.

We fit the spectra using the {\sc XSPEC} package (version 11.3.0;
\citealt{1996adass...5...17A}). We fit the spectrum with an absorbed
black body model and with an absorbed neutron star atmosphere model
(NSA; \citealt{1991MNRAS.253..193P}; \citealt{1996A&A...315..141Z}). We
kept the absorption fixed at the value we found during outburst (${\rm
N_H=4.4\times10^{21} cm^{-2}}$; \citealt{2003MNRAS.341..823J}). The
Galactic absorption, the NSA normalisation, the mass, and the radius
of the neutron star, were held fixed during the fit at
$4.4\times10^{21}$ cm$^{-2}$, $\frac{1}{D^2}=1.3\times10^{-8}$
pc$^{-2}$ (for the distance, D in pc, we took the value of the
Globular Cluster 8.8$\times10^3$ pc, see Section 2.2), and 1.4
M$_\odot$, 10 km, respectively. The best--fitting parameters are
presented in Table~\ref{fitpars}. The reddening to the Globular
Cluster NGC~6293 of ${\rm E(B-V)=0.41}$ implies a column density of
${\rm 2.2\times10^{21} cm^{-2}}$ assuming ${\rm A_V = 3.1 \times
E(B-V)}$ and using the relation between ${\rm A_V}$ and ${\rm N_H}$ of
\citet{1995A&A...293..889P}. We also fitted a NSA and a black body to
the data with the ${\rm N_H}$ fixed at ${\rm 2.2\times10^{21}
cm^{-2}}$ (see Table~\ref{fitpars}). The best--fitting parameters are
consistent within the 90 per cent confidence contours with those
obtained using ${\rm N_H=4.4\times10^{21} cm^{-2}}$.

\begin{figure*}
  \includegraphics[angle=-90,width=13cm]{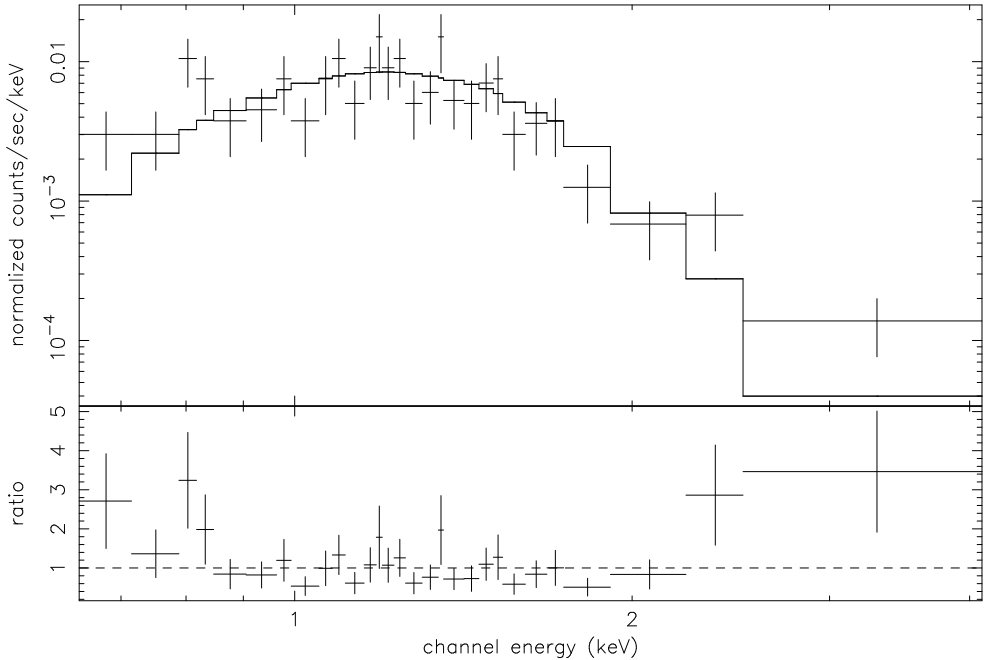}
\caption{{\it Top panel}: The quiescent X--ray spectrum of
  XTE~J1709--267 detected with {\it Chandra} fitted with an absorbed
  black body model. {\it Bottom panel}: the ratio between the data and
  the best--fitting model.}
\label{bbfit}
\end{figure*}

The flux in the last two bins is underestimated for both the black body (see
Figure~\ref{bbfit}) and the NSA model fits. This can (partially) be explained by
the fact that we did not subtract the background. Had we subtracted the
background as defined above it would have reduced the count rate above 3 keV (3
out of the 10 photons detected above 3 keV would have been labelled background
photons). Additionally, a hard (power--law) spectral could be present. A
power--law spectral component with index 2 contributes less than 20 per cent to
the unabsorbed flux in the 0.5--10 keV band. The absorbed 0.5--10 keV source
flux for both models and both ${\rm N_H}$ values considered above was consistent
with $\sim(4.6\pm0.4)\times10^{-14}$ erg cm$^{-2}$ s$^{-1}$, whereas the
unabsorbed flux was $\sim(1.0\pm0.3)\times10^{-13}$ erg cm$^{-2}$ s$^{-1}$. Here
the error is determined from the range in fluxes derived from the various
models. For all models we performed a Monte Carlo simulation (using the {\sc
goodness} command in {\sc XSPEC}). We simulated 10$^4$ spectra based on a
Gaussian distribution of parameters centred on the best--fit model parameters
with a Gaussian width set by the 1$\sigma$ errors on the fit parameters.  The
percentage of these simulations with the fit statistic less than that for the
data is more than the fiducial 50 per cent mark for 3 of the 4 cases considered,
but again this may be partially explained by the presence of background photons
in the last two bins. The goodness percentages are given in Table~\ref{fitpars}.

\begin{table*}
\caption{Best fit parameters of the quiescent spectrum of
XTE~J1709--267 (NSA stands for neutron star atmosphere and BB refers
to black body).  All quoted errors are at the 68 per cent confidence
level. }

\label{fitpars}
\begin{center}

\begin{tabular}{lcccc}
\hline 
${\rm N_H}^a$ & Model & BB radius & Temperature & Goodness \\ 
cm$^{-2}$ & & in ${\rm (\frac{d}{10\,kpc})^2}$ km & keV &  per cent\\
\hline 
\hline
4.4$\times10^{21}$& BB       & 3.6$\pm$1.2 & 0.24$\pm0.02$ & 85 \\
4.4$\times10^{21}$& NSA      & --- &  0.125$\pm0.002$& 57 \\
2.2$\times10^{21}$& BB       & 1.2$\pm$0.4 & 0.28$\pm0.02$ & 49 \\
2.2$\times10^{21}$& NSA      & --- &  0.116$\pm0.002$& 82 \\

\end{tabular}
\end{center}
{\footnotesize $^a$ Parameter fixed at this value. }\newline
\end{table*}

Even though the fit results presented in \citet{2003MNRAS.341..823J} are still
valid, we also refitted the spectra of the last 3 of the 5 2002 {\it Chandra}
observations of XTE~J1709--267, i.e.~those with IDs 3464, 3475, 3492 presented
by \citet{2003MNRAS.341..823J}. We used a fit--function consisting of a
black--body plus a power--law component. For this we reprocessed the archival
data with the newest calibration files available. The result of these spectral
fits can be found in Table~\ref{refit}. These spectra were obtained while the
source returned to quiescence after an outburst.

\begin{table*} \caption{Results of fits to spectra from XTE~J1709--267 observations
obtained immediately after an outburst in 2002 (observations with IDs 3464, 3475, and 3492)
using a black body plus power law spectral model (see also Jonker et al.~2003). All quoted
errors are at the 68 per cent confidence level. The ${\rm N_H}$ was kept fixed at a value of
4.4$\times10^{21}$ cm$^{-2}$ during the fits. On the last line we give the result for a
spectral fit using to the combined data from observation ID 3492 \& 3507.}

\label{refit}
\begin{center}

\begin{tabular}{cccccccc}
\hline 
Obs. & Date & BB$^a$ radius & BB temp. & PL$^a$ index & PL norm.$^b$ & $\chi^2$/d.o.f. /  & Unabs.~flux (0.5-10 keV)\\ 
ID & MJD (UTC) & in ${\rm (\frac{d}{10\,kpc})^2}$ km & keV &  & photons keV$^{-1}$~cm$^{-2}$~s$^{-1}$  & Goodness & erg cm$^{-2}$ s$^{-1}$\\
\hline 
\hline
3464 & 52365.018 & 60$\pm$30 & 0.22$\pm$0.02 & 2.5$\pm0.2$ & $(9\pm2)\times10^{-4}$ & 98/111 & 4.7$\times10^{-12}$\\
3475 & 52374.727 & (3$^{+3}_{-2}$)$\times10^2$ & 0.13$\pm$0.02 & 2$^c$ & $(8\pm2)\times10^{-5}$ & 23/24 & 8.4$\times10^{-13}$\\
3492 & 52387.898 & 6$\pm$4  & 0.26$\pm$0.03 & .... & .... & 75\% & 2.4$\times10^{-13}$\\
3492 \& 3507& 52771.322 & 7$\pm$3  & 0.22$\pm$0.02& .... & .... & 60\% & 1.7$\times10^{-13}$\\
\end{tabular}
\end{center}
{\footnotesize $^a$ BB = black body, PL = power law. }\newline
{\footnotesize $^b$ Power law normalisation at 1 keV.}\newline
{\footnotesize $^c$ Power law index fixed at this value.}\newline
\end{table*}

\subsection{RXTE}

\begin{figure*}
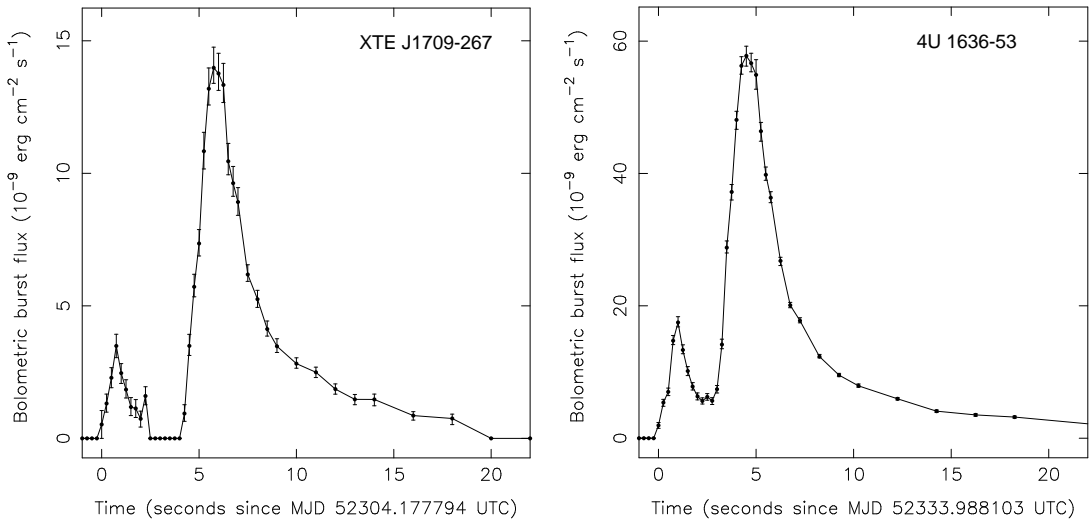

  \includegraphics[width=7cm]{burst.lc.flux.ps}
\quad
\includegraphics[width=7cm]{1636.burst.pre.cursor.ps}
\caption{{\it Left panel:} The type~I X--ray burst detected in
    XTE~J1709--267 with {\it RXTE}'s PCA on MJD 52304.177794
    (UTC). Note the presence of a precursor event. The zeros in
    between the precursor and the main peak and before the precursor
    are the result of the subtraction of the persistent non--burst
    emission. {\it Right panel:} A type~I X--ray burst with a similar
    precursor event detected in 4U~1636--53 with {\it RXTE}'s PCA on
    MJD 52333.988103. In both cases PCUs 0234 were operational.}
\label{burst.lc}
\end{figure*}

We found three type~I X--ray bursts in the {\it RXTE} proportional counter
array (PCA) observations obtained during the outburst of the source in 2002. We
analysed these bursts using {\sc ftools 5.2}; we present the results for the
burst for which 4 of the 5 Proportional Counter Units (PCUs) were operational
(for the other two bursts only three PCUs were operational; the profile of
these two bursts was similar to that of the burst discussed in detail here).
The burst start--time is MJD 52304.177794(5) UTC; the last digit in between
brackets denotes the uncertainty. The lightcurve of the burst is plotted in
Figure~\ref{burst.lc} ({\it left panel}). A precursor to the main burst event
can clearly be seen. No burst oscillations were present in the range 50--2000
Hz in the 2.5--25 keV power spectrum with a 95 per cent confidence upper limit of
14 per cent.

\begin{figure*}
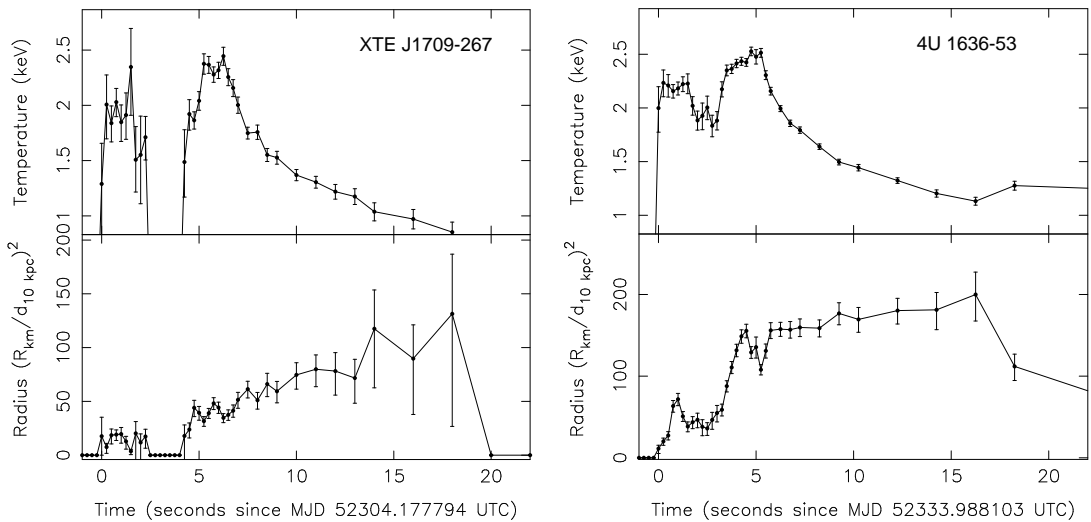
 \includegraphics[width=7cm]{1709.rad.temp.ps} \quad
\includegraphics[width=7cm]{1636.rad.temp.ps} \caption{{\it Left panel:} The evolution of the black
body temperature ({\it top}) and radius ({\it bottom}) as a function of time since the burst start as
determined from X--ray spectral fits of {\it RXTE} PCA data in the energy range from 2.5--25 keV for
XTE~J1709--267. {\it Right panel:} The same for the burst of 4U~1636--53. } \label{burstspec} \end{figure*}

Spectra of the burst in the energy range 2.5--25 keV were
calculated. We subtracted the persistent emission averaged over 16
seconds starting 20 seconds before the burst onset.  We fitted the
resultant burst spectra with an absorbed black body. The absorption
was fixed at ${\rm N_H}\sim2\times10^{21}$ cm$^{-2}$, this is
consistent with the value derived for the Globular Cluster although
our results were consistent with being the same when we used ${\rm
N_H}\sim4.4\times10^{21}$ cm$^{-2}$; the {\it RXTE} data is not very
sensitive to the exact value for ${\rm N_H}$ as long as it is lower
than a few times $10^{22}$ cm$^{-2}$. The resultant black body radius
and temperature for the burst spectra are presented in
Figure~\ref{burstspec} ({\it left panel}). The bolometric peak flux
($\sigma T^4 (\frac{R}{d})^2$) we found was
$(1.40\pm0.06)\times10^{-8} {\rm erg\, cm^{-2}\, s^{-1}}$.

Type~I X--ray bursts with a precursor event have previously been
identified with extreme cases of photospheric radius expansion, where
the photospheric radius becomes so large and the black body
temperature so small that the peak of the emission drops below the
X--ray band (\citealt{levata1993}). However, from our spectral study of
the burst we do not find evidence for a large increase in black body
radius as is found for other photospheric radius expansion bursts with
precursors. To investigate further whether this burst is in fact a
photospheric radius expansion burst we compare the burst properties
with that of a nearly identical burst we found sifting through
archival {\it RXTE} observations of the source 4U~1636--53. The
findings for 4U~1636--53 were plotted in the {\it right panels} of
Figures~\ref{burst.lc} and ~\ref{burstspec}. The two bursts are
remarkably similar to each other. However, for 4U~1636--53
photospheric radius expansion bursts are observed relatively often
(see \citealt{1988A&A...199L...9F}). The mean peak flux for the bright
photospheric radius expansion bursts for 4U~1636--53 is
$(7.2\pm0.7)\times10^{-8} {\rm erg\,cm^{-2}\,s^{-1}}$ (here the
uncertainty is the standard deviation of the observed peak
photospheric radius expansion burst flux distribution; Galloway et
al.~in preparation). With a value of $(5.8\pm0.1)\times10^{-8} {\rm
erg\,cm^{-2}\,s^{-1}}$ the bolometric peak flux of the burst with
precursor in 4U~1636--53 is significantly lower.

The burst from 4U~1636--53 reached a peak flux a factor of 1.25
smaller than the mean peak flux of the photospheric radius--expansion
bursts from that source. If we identify the latter as the Eddington
limit, and multiply the peak flux of the burst from XTE~J1709--267 by
this factor we infer a distance of
10.1$^{+3.2}_{-1.7}$/13.9$^{+4.4}_{-2.3}$ kpc for the source for an
assumed Eddington limit of 2.0/3.8$\times10^{38}{\rm erg\,s^{-1}}$
(see \citealt{2003A&A...399..663K}). The two values correspond
approximately to hydrogen rich and hydrogen poor bursts,
respectively. Of these distances only the lower value is consistent
with the suggestion made previously (\citealt{2001A&A...368..137V};
\citealt{2003MNRAS.341..823J}) that XTE~J1709--267 is associated with
the metal poor Globular Cluster NGC~6293, since the distance to
NGC~6293 is known to be $\sim8.8$ kpc (see
\citealt{1996AJ....112.1487H}; according to \citealt{1991AJ....101.2097J}
the uncertainties in the cluster distance are considerable). If, as
has been argued by several authors (cf.~\citealt{1999ApJ...512..892T};
\citealt{baolbo2000}; \citealt{2003A&A...399..663K}), the flux reported
from measurements made with the {\it RXTE} satellite are
systematically too high by approximately 20 per cent then the two
distances we quote above are too small by a factor $\sim1.12$.

\subsection{Optical observations}

\begin{figure*}
  \includegraphics[width=15cm]{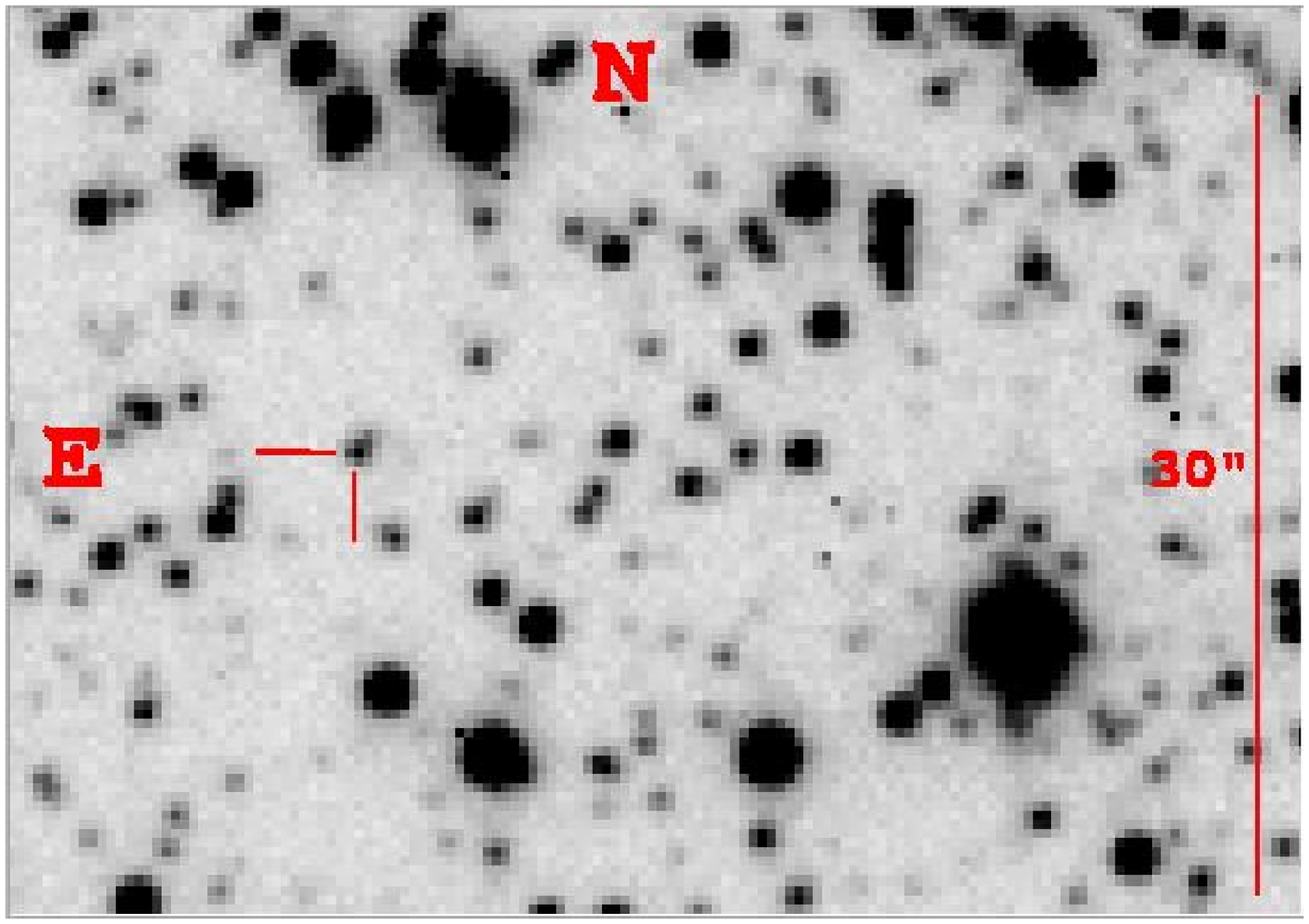}
\quad
\includegraphics[width=15cm]{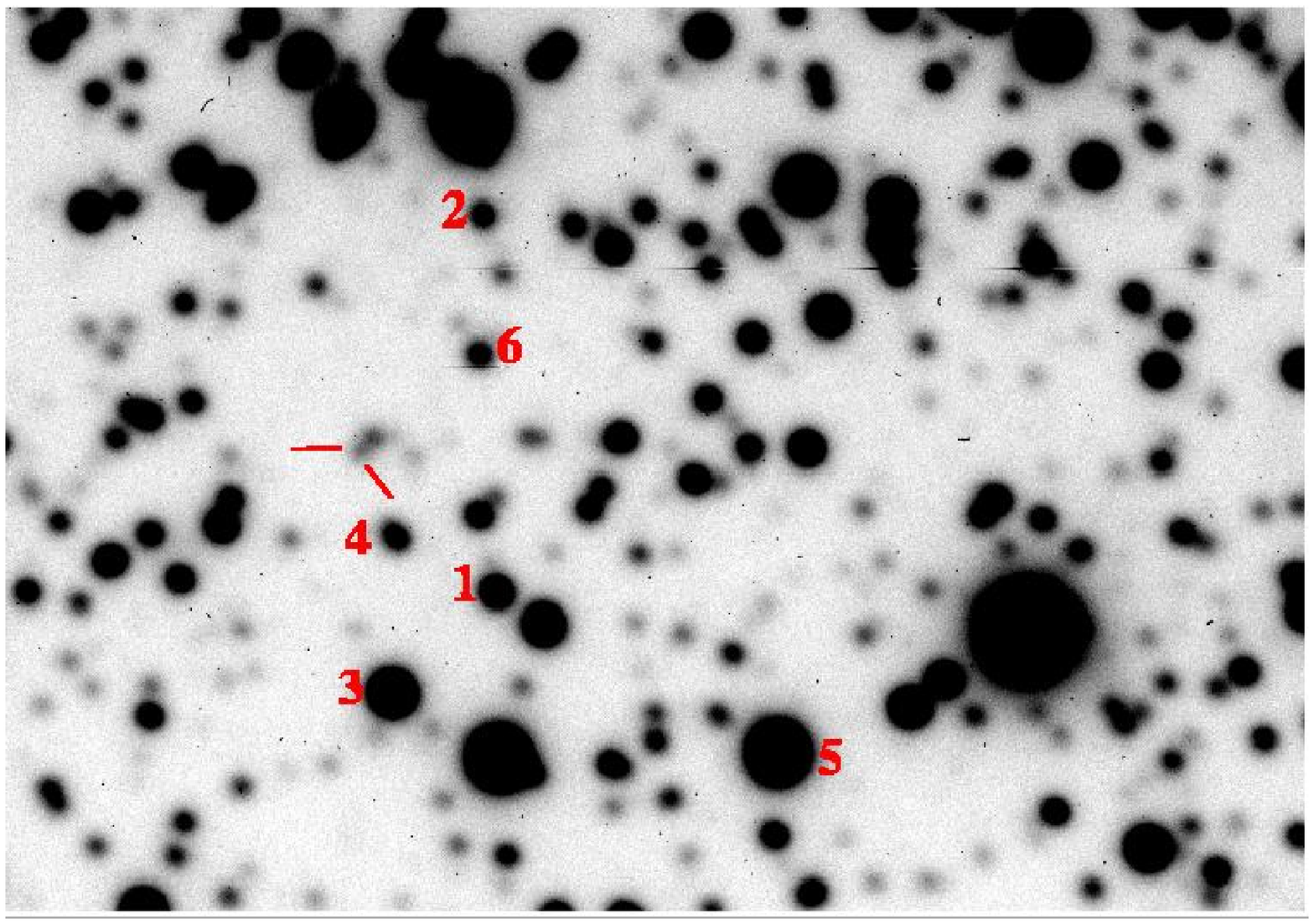}
\caption{ {\it Top panel}: The 900~s $R$--band image obtained on March
26, 2004 with the ANDICAM instrument mounted on the 1.3~m telescope at
CTIO. The optical counterpart to the neutron star SXT XTE~J1709--267
is indicated with tick--marks. The pixel scale is 0.369 arc seconds
per pixel. {\it Bottom panel}: The 20 minute $R$--band image obtained
on June 1, 2003 with the MAGIC instrument on the 6.5~m Magellan Clay
telescope at Las Campanas. The photometric reference stars have been
indicated with a number. The proposed counterpart was detected in
quiescence at $R\sim$22.2. The image has been rebinned to a pixel
scale of 0.138 arc seconds per pixel. }
\label{optical}
\end{figure*}

We observed the field of XTE~J1709--267 for 900~s through an $R$--band filter
with ANDICAM mounted on the 1.3 m telescope at CTIO on March 26, 2004 (MJD
53090.330498 UTC). The seeing during the observations was approximately 0.9 arc
seconds. The image was corrected for bias and pixel--to--pixel variations in
sensitivity. An X--ray outburst was reported to have started not long before
MJD 53070 UTC. The X--ray flux on Mar 19.83 (MJD 53083.83 UTC,
i.e.~approximately seven days before the optical observations) was reported to
be 32 mCrab (\citealt{2004ATel..255....1M}). Therefore, it is likely that on
March 26, 2004 the source was still active in X--rays. We found a source
located in the 0.5 arc second {\it Chandra} error circle which was not detected
with a limiting magnitude of 20.5 on a $V$--band image obtained with the 2.5~m
Isaac Newton Telescope located on La Palma on June 6, 2002 during quiescence
(see \citealt{2003MNRAS.341..823J}). We conclude that we discovered the
counterpart to the neutron star SXT XTE~J1709--267 (\citealt{ateljonker}; see
Figure~\ref{optical} {\it top panel}).  A nearby interloper to the North--West
of the counterpart at an angular distance of $\sim$0.7 arc seconds can be seen.
Due to the large pixel scale of ANDICAM and the seeing of 0.9 arc seconds this
star was not well separated in the ANDICAM image. The night was photometric and
the $R$--band magnitude of the counterpart plus the nearby interloper combined
was found to be 20.15$\pm$0.04.

We also obtained two images with an exposure time of 10 minutes each through an
$R$--band filter with MAGIC mounted on the 6.5~m Magellan Clay telescope at Las
Campanas on June 1, 2003 (52791.360822 UTC) when the source was in quiescence.
The seeing during these observations was approximately 0.9 arc seconds. The
images were corrected for bias and pixel--to--pixel variations in sensitivity.
Further, we combined the two 10 minute images. We used photometry of 6 stars
close to the position of XTE~J1709--267 relative to the ANDICAM image to
transform the observed count rates to $R$--band magnitudes. These reference
stars are indicated with the numbers 1--6 in the {\it bottom panel} of
Figure~\ref{optical}. If we assume that the magnitude of the nearby interloper
did not change between the image obtained in outburst and in quiescence it
contributed $\sim$25 per cent to the outburst magnitude mentioned above. If we
correct for this the outburst $R$--band magnitude of the counterpart becomes
$R$=20.5 with an estimated error of 0.1. The quiescent counterpart $R$--band
magnitude is $R$=22.24$\pm$0.03. The magnitudes of the 6 photometric reference
stars, the magnitude of the proposed counterpart in quiescence and outburst,
and the magnitude of the star close to the position of the counterpart, derived
using point--spread function fitting routines of the {\sl DAOPHOT II} package
(\citealt{1987PASP...99..191S}) in \textsc {iraf} \footnote{\textsc {iraf} is
distributed by the National Optical Astronomy Observatories} are given in
Table~\ref{mags}. We obtained an astrometric solution with an rms error of 0.3
arcseconds for the combined image using the known positions of 9 nearby USNO
B1.0 stars which were relatively well separated from other stars; these USNO
B1.0 stars have identification numbers 0633--0521300, 0521161, 0520994,
0520833, 0520757, 0520873, 0520981, 0521064, and 0521091. The position of the
proposed counterpart is consistent with that of the X--ray transient as
determined with {\it Chandra} (\citealt{2003MNRAS.341..823J}). The nearby
interloper position lies just outside the {\it Chandra} error circle

\begin{table}
\caption{The magnitudes of 6 comparison stars numbered as
indicated in the {\it bottom panel} of Figure~\ref{optical} which have
been used as photometric calibrators, the proposed counterpart (CP)
and that of the star near (within 0.7 arc seconds) the position of the
proposed counterpart. The errors are formal fitting errors only; the
estimated error in the zeropoint is $\sim$0.1 magnitude.}

\label{mags}
\begin{center}

\begin{tabular}{cc}
\hline 
Star & $R$--band magnitude \\
\hline 
\hline
1    &  19.41$\pm$0.02 \\
2    &  20.26$\pm$0.05 \\
3    &  17.82$\pm$0.01 \\
4    &  20.35$\pm$0.04  \\
5    &  16.57$\pm$0.01 \\
6    &  20.47$\pm$0.04 \\
CP    & 20.5$\pm$0.1$^a$/22.24$\pm$0.03$^b$   \\
Nearby star & 21.56$\pm$0.02 \\
\end{tabular}
\end{center}
{\footnotesize $^a$ Value in outburst, corrected for the contribution
of the nearby interloper}\newline {\footnotesize $^b$ Value in
quiescence}\newline
\end{table}

\section{Discussion}

We observed the neutron star SXT XTE~J1709--267 on several occasions
in outburst and quiescence in X--ray and optical.  The
brightening of a star from $R$=22.24$\pm$0.03 in quiescence to
$R$=20.5$\pm$0.1 in outburst and its positional coincidence with the
Chandra position for XTE~J1709--267 provides convincing evidence that
we have discovered the optical counterpart
(\citealt{ateljonker}). Obviously, there is a small chance that the star
found in quiescence is an interloper and not the true counterpart;
spectroscopic observations of the proposed quiescent counterpart will
test this. If we assume, however, that we also detected the optical
counterpart in quiescence, the absolute magnitude, ${\rm M_R}$, for
the counterpart would be ${\rm M_R}=6.6-5.7$ where we have assumed an
$R$--band interstellar absorption, ${\rm A_R}$, of 0.92--1.84
magnitudes, respectively [for a distance of 8.8 kpc, from ${\rm
N_H=(2.2-4.4)\times10^{21} cm^{-2}}$, and from the relations between
${\rm N_H}$, ${\rm A_V}$, and ${\rm A_R}$ given by
\citealt{1985ApJ...288..618R} and \citealt{1995A&A...293..889P}]. These
${\rm M_R}$--values are consistent with XTE~J1709--267 being a
low--mass X--ray binary with a late type (K) dwarf companion.

The unabsorbed luminosity in the 0.5--10 keV X--ray band we find using {\it
Chandra} observations of the source in quiescence is only slightly lower than
the lowest luminosity measured by \citet{2003MNRAS.341..823J} (approximately a
factor 2; in that paper we gave the unabsorbed flux in the 0.1--10 keV and not
the 0.5--10 keV band). So, the decay in luminosity in about 14 months is small
(approximately factor of 2). This is in contrast with the decay rate of the
quasi--persistent source MXB~1659--29. \citet{2004wijninpress} found that for
that source the bolometric luminosity decreased by a factor 7--9 in 18 months.
Such a difference could be explained by the fact that MXB~1659--29 had been
accreting steadily for several years before returning to quiescence. This
extended period of steady accretion may have heated the neutron star crust to
temperatures higher than that of the neutron star core and after the outburst
the crust cools down (\citealt{2004wijninpress}). However, this difference
could also reflect a difference between the quiescent mass accretion rates,
although it is unclear why in some sources the mass accretion rate hits a
minimum close after the outburst whereas in other neutron star STXs like
MXB~1659--29 the mass accretion rate keeps decreasing gradually.

\begin{figure*}  \includegraphics[angle=0,width=14cm]{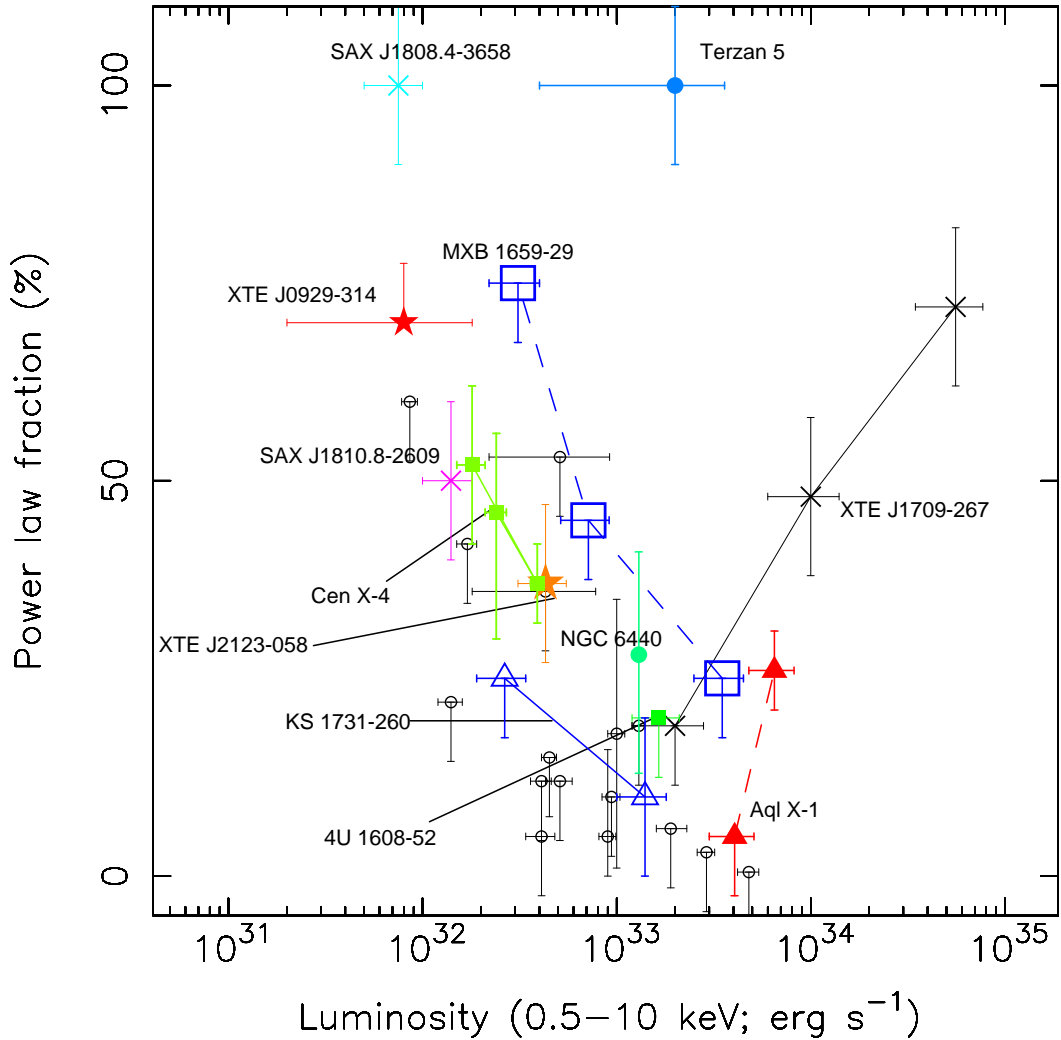} 
\caption{The fractional power--law contribution to the unabsorbed quiescent
0.5--10 keV luminosity for quiescent neutron stars. The points without positive
error bars denote upper limits on the power--law fraction, for XTE~J0929--314
there is a lower limit to the power--law contribution of 70 per cent. Data
points for different sources are indicated with different symbols and/or
colours. For Aql~X--1, Cen~X--4, MBX~1659--29, XTE~J1709--267, and KS~1731--260
we have plotted more than one point since the quiescent source luminosity and
power--law fraction of the quiescent luminosity was found to vary (for
references see text). These multiple data points of the same source have been
connected with a (dashed) line. The Galactic Cluster sources, selected by
Heinke et al.~(2003) on the basis of their soft spectrum (the power law
contribution had to be less than 40 per cent for the sources to be selected)
are indicated with a small circle. } \label{relation} \end{figure*}

Recently, we found (\citealt{2004MNRAS.349...94J}) that there seems to be an
anti--correlation between the fractional power--law contribution to the 0.5--10
keV luminosity and the source luminosity in quiescence for quiescent
luminosities lower than $\sim 1 - 2\times10^{33}$ erg s$^{-1}$ and a
correlation between these two parameters for luminosities above this
luminosity. In Figure~\ref{relation} we plot the power--law fractional
contribution to the total 0.5--10 keV unabsorbed quiescent luminosity for
several neutron star SXTs for which the distance is accurately known\footnote{
Data for Cen~X--4 was taken from \citet{1996PASJ...48..257A},
\citet{2001ApJ...551..921R}, and \citet{2004ApJ...601..474C}, for 4U~1608--52
from \citet{1996PASJ...48..257A}, for MXB~1659--29 from
\citet{2004wijninpress}, for XTE~J1709--267 from this work, for KS~1731--260
from \citet{2001ApJ...560L.159W}, \citet{2002ApJ...577..346R}, and
\citet{2002ApJ...573L..45W}, for Terzan~5 from \citet{wijnsub2004}, for
NGC~6440 from \citet{2001ApJ...563L..41I} and \citet{2003ApJ...598..501H}, for
SAX~J1808.4--3658 from \citet{2002ApJ...575L..15C}, for SAX~J1810.8--2609 from
\citet{2004MNRAS.349...94J}, for Aql~X--1 from \citet{2002ApJ...577..346R}, 
for XTE~J2123--058 from \citet{TOMSICK2123} and finally, for XTE~J0929--314
from \citet{wijnsub2.2004}.}. To obtain the quiescent luminosities we use
distances quoted in \citet{peterandgijssubm} for the systems where photospheric
radius expansion bursts have been observed. Since the photospheric radius
expansion burst luminosity is thought to be close to 2.0 or 3.8$\times10^{38}
{\rm erg\, s^{-1}}$ (see \citealt{2003A&A...399..663K}) we use a range in
quiescent luminosities to account for this ambiguity in source distance. In the
case of XTE~J1709--267 we took a distance range of 8--12 kpc. For the neutron
star SXTs in the Globular Clusters Terzan~5 and NGC~6440 we took a distance of
8.7$\pm3$ kpc and 8.5$\pm0.4$ kpc, respectively (\citealt{2002ApJ...571..818C};
\citealt{1994A&AS..108..653O}). Finally for Cen~X--4 we use a distance of 1.2
kpc (\citealt{1980ApJ...241..779K}; \citealt{1989A&A...210..114C};
\citealt{1996ApJ...473..963B}). We do not take into account errors on the
source quiescent luminosities due to errors on the measured source flux since
these are typically smaller than the uncertainty in the burst chemical
composition. When the error on the power--law contribution to the flux was not
given in the literature we assumed an error of 10 per cent. Finally, we
included in the plot Globular Cluster sources which are thought to be quiescent
neutron star SXTs on the basis of their soft spectrum
(e.g.~\citealt{1984MNRAS.210..899V}). We used the sources and limits on the
power law component in the spectrum as found by \citet{2003ApJ...598..501H}.
\footnote{We took a distance of 5.3$\pm$0.3 kpc for the Globular Cluster
$\Omega$~Cen (\citealt{2001AJ....121.3089T}), 10.3$\pm$0.8 kpc for M~80
(\citealt{1996A&A...311..778B}), 5.2$\pm$0.3 kpc for 47~Tuc, 3.6$\pm$0.3 kpc
for NGC~6397, and 9.5$\pm$0.9 kpc for M~30 (all from
\citealt{2000ApJ...533..215C}).}

To the high luminosity side, the trend of increasing power--law fraction with luminosity is
dominated by the data points of XTE~J1709--267 which was followed by {\it Chandra} during
its decay to quiescence after an outburst (Jonker et al.~2003; see also Table~\ref{refit}).
We found that the power law contribution to the 0.5--10 keV X--ray luminosity decreased from
72 per cent on MJD 52365.018, to 48 per cent on MJD 52374.727, to less than 19 per cent
using the combined data from observation 3492 and 3507. The detailed study of Aql~X--1
confirms the observed trend (Rutledge et al.~2002). However, the quiescent properties of the
counterpart to the neutron star SXT EXO~1745--248 in the dense Globular Cluster core of
Terzan~5 seemingly do not fit the correlation (Wijnands et al.~2004). Perhaps the identified
source is an interloper or perhaps the distance to Terzan~5 is much smaller than what is
assumed (\citealt{1996A&A...308..733O} derived a distance of 5.6 kpc). Alternatively, the
apparent correlation is spurious and Terzan~5 is the first source to fill--in the gap. If
the apparent smooth change in power--law contribution to the quiescent luminosity is real it
could mean that the nature of the power--law spectral component at high and low source
luminosities is different.

Since we observed the power--law contribution in XTE~J1709--267 to decrease
when the source returned to quiescence after an outburst it is conceivable that
the power--law component at luminosities above $\sim 1 - 2\times10^{33}$ erg
s$^{-1}$ finds its origin in residual accretion. It has been proposed that
neutron star SXTs enter a propeller regime when the outburst decay rate
steepens impeding most if not all accretion (\citealt{1998ApJ...499L..65C}).
However, pulsations were still detected in SAX~J1808.4--3658 after the alleged
onset of the propeller mechanism (\citealt{1999ApJ...521..332P}). Furthermore,
steepening of the decay is also found in black hole candidate SXTs
(e.g.~\citealt{2004MNRAS.inpress}). Finally, the work of
\citet{2002A&A...392..931C} shows that there is a class of burst sources which
likely accrete at a low level. Deep observations a few hours after the
detection of a type~I burst in SAX~J2224+5421 did not reveal a persistent
source with a 2--10 keV upper limit of 1.3$\times10^{-13}$ erg cm$^{-2}$
s$^{-1}$ (\citealt{2003A&A...405.1033C}), which for the distance of
SAX~J2224+5421 leads to an upper limit on the luminosity in that band of
7.4$\times10^{32}$ erg s$^{-1}$. Therefore, we think it is more likely that the
origin of the power--law at relatively high quiescent luminosities lies in
residual accretion.  If so this can help explain the observed short term
neutron star temperature changes in Aql~X--1 (\citealt{2002ApJ...577..346R}).
These short term temperature changes pose a problem for the cooling neutron
star core/crust model. However, if residual accretion is ongoing the observed
changes can, for instance, be explained as being the result of a different
hydrogen and helium content in the atmosphere during the two observations
caused by the residual accretion. Hence, the fact that this power--law
component can be explained at least qualitatively in terms of residual
accretion helps the cooling neutron star model as an explanation for the
thermal component.

The origin of the power--law on the low quiescent luminosities side of $\sim 1 -
2\times10^{33}$ erg s$^{-1}$ is still unclear. Since SAX~J1808.4--3658 is known to have
a sizable magnetic field the power--law component could be explained as being due to a
pulsar--wind mechanism (e.g.~\citealt{1994ApJ...423L..47S};
\citealt{2003A&A...404L..43B}). However, this cannot explain the strong power--law
components in the non--pulsating sources Cen~X--4, SAX~J1810.8--2609, and XTE~J2123--058
although one must bear in mind that Cen~X--4 and SAX~J1810.8--2609 have not been
observed with {\it RXTE}/PCA when the sources were in outburst (whereas XTE~J2123--058
{\it has} been observed with {\it RXTE}/PCA in outburst see
\citealt{1999ApJ...513L.119H} and \citealt{1999ApJ...521..341T}).  A possible
explanation for the apparent correlation between the fractional power--law contribution
and the total 0.5--10 keV luminosity at low quiescent luminosities is that there is a
power--law spectral component with a luminosity of $\sim10^{32}\,{\rm erg\,s^{-1}}$.
This power--law luminosity would then need to be approximately the same for all sources;
the luminosity of the black body can differ between sources. In the cooling neutron star
model the luminosity of the thermal component depends on the time averaged mass
accretion rate and the neutron star mass (\citealt{1998ApJ...504L..95B};
\citealt{2001ApJ...548L.175C}). A low black body luminosity could point at a low
time--averaged mass accretion rate and/or a large neutron star mass allowing for
enhanced core cooling (cf.~\citealt{2001ApJ...548L.175C}).  The nature of the power--law
and why it would have a luminosity close to $\sim10^{32}\,{\rm erg\,s^{-1}}$ in the
0.5--10 keV band is unclear.

Finally, we found a precursor to several type~I X--ray bursts of
XTE~J1709--267. A similar burst with a precursor was found for the bursting
atoll source 4U~1636--53. A precursor to the main burst event in relatively
long bursts has been associated with photospheric radius expansion bursts (see
\citealt{levata1993}).  However, from a comparison of the burst properties of
photospheric radius expansion bursts in 4U~1636--53 (Galloway et al.~in
preparation) with the properties of the burst with a precursor in 4U~1636--53
we conclude that these bursts with precursors in 4U~1636--53 and hence also in
XTE~J1709--267 are not photospheric radius expansion bursts. Perhaps these
precursor events are related to bursts with multiple peaks observed in
4U~1636--53 (\citealt{1986MNRAS.221..617V}; \citealt{1988A&A...199L...9F}).

\section*{Acknowledgments}  \noindent  Support for this work was provided by NASA
through Chandra Postdoctoral Fellowship grant number PF3--40027 awarded by the Chandra
X--ray Center, which is operated by the Smithsonian Astrophysical Observatory for NASA
under contract NAS8--39073. MRG acknowledges support from LTSA Grant NAG5--10889 and
Contract NAS8--39073 to the Chandra X--ray Center. PGJ would like to thank Sergio
Campana and Albert Kong for comments on an earlier version of the manuscript.

\end{document}